\documentclass[twocolumn]{aastex62}

\newcommand{\drw}{DR21W}
\newcommand{\Blos}{\ensuremath{B_{\text{los}}}}
\newcommand{\meth}{CH$_3$OH}

\newcommand{\kmS}{km~s$^{-1}$}

\usepackage{ulem} 

\shorttitle{44 GHz Methanol Maser Zeeman Effect toward DR21W}
\shortauthors{Momjian \& Sarma}

\accepted{\textit{2018 December 22}}
\reportnum{\textit{To be published in}}
\reportnum{\textit{The Astrophysical Journal}}

\begin{document}

\title{THE ZEEMAN EFFECT IN THE 44 GHz \\ CLASS I METHANOL (CH$_3$OH) MASER LINE TOWARD DR21W}

\correspondingauthor{Emmanuel Momjian}
\email{emomjian@nrao.edu}

\author{E.~Momjian}
\affiliation{National Radio Astronomy Observatory, P.O. Box O, Socorro, NM 87801, USA, emomjian@nrao.edu}

\author{A.~P.~Sarma}
\affiliation{Physics Department, DePaul University, 2219 N.~Kenmore Ave., Chicago, IL 60614,
USA; asarma@depaul.edu}

\begin{abstract}
We report the detection of the Zeeman effect in the 44 GHz Class I methanol maser line toward the high mass star forming region DR21W. There are two prominent maser spots in DR21W at the ends of a northwest-southeast linear arrangement. For the maser at the northwestern end (maser A), we fit three Gaussian components. In the strongest component, we obtain a significant Zeeman  detection, with $z\Blos=-23.4\pm3.2$~Hz. If we use $z=-0.920$~Hz~mG$^{-1}$ for the $F=5 \rightarrow 4$ hyperfine transition, this corresponds to a magnetic field $|\Blos|=25.4$~mG; \Blos\ would be higher if a different hyperfine was responsible for the 44 GHz maser, but our results also rule out some hyperfines, since fields in these regions cannot be hundreds of mG. Class I methanol masers form in outflows where shocks compress magnetic fields in proportion to gas density. Designating our detected $\Blos=25$~mG as the magnetic field in the post-shock gas, we find that $\Blos$ in the pre-shock gas should be 0.1--0.8 mG. Although there are no thermal-line Zeeman detections toward DR21W, such values are in good agreement with Zeeman measurements in the CN thermal line of 0.36 and 0.71~mG about $3.5\arcmin$ away in DR21(OH) in gas of comparable density to the pre-shock gas density in DR21W. Comparison of our derived magnetic energy density to the kinetic energy density in DR21W indicates that magnetic fields likely play a significant role in shaping the dynamics of the post-shocked gas in DR21W. 

\end{abstract}


\section{Introduction}

Observations of massive star forming regions require high spatial resolution because higher mass stars form in densely populated environments (\citealt{motte2017}). Being point sources with high intensity, masers offer us the opportunity to study such regions at extremely high spatial resolution (see, e.g., \citealt{sanna2017}; \citealt{mosca2016}). Class I methanol (\meth) masers usually form in outflows where collisional shocks create enhanced \meth\ abundances and pump the maser transitions \citep[see, e.g.,][]{leurini+2016}. As part of a long-term effort to measure magnetic fields in regions traced by \meth\ masers, we present a detection of the Zeeman effect in the 44 GHz Class I \meth\ maser line toward the high mass star forming region DR21W. Magnetic fields are known to play an important role in star formation (e.g., \citealt{huabai2014}), and measurements such as those reported here add to a compendium of magnetic fields in a range of environments that will allow us to understand in detail their role before and during the star formation process. 

DR21W is part of the high mass star forming complex DR21, located at a distance of $1.5^{+0.08}_{-0.07}$~kpc (\citealt{rygl2012}).  In a velocity-integrated C$^{18}$O map by \citet{wm1990}, the DR21/W75 complex presents itself as a column of molecular emission extending $\sim$17\arcmin\ along the north-south direction, with DR21 in the southern portion. Line wings in the C$^{18}$O spectra, together with low intensity C$^{18}$O emission, delineate an outflow in DR21 running roughly from north-of-east to south-of-west that is also seen in vibrationally excited H$_2$ emission (\citealt{davis2007}). It is in the prominent western lobe of this outflow that the 44 GHz Class I \meth\ masers of DR21W are located (\citealt{kogan1998}). \citet{pm1990} observed a Class I \meth\ maser at 95 GHz toward DR21W, located between a cluster of CS emission peaks and a broad ridge of vibrationally excited H$_2$ emission. The CS emission peaks trace ambient dense molecular gas, whereas the H$_2$ emission traces the shock fronts associated with outflows. \citet{lw1997} observed non-masing emission in four $2_k \rightarrow 1_k$ \meth\ lines with the IRAM Plateau de Bure interferometer, and found this thermal emission to be distributed in several clumps in an elongated structure, oriented northeast to southwest, that surrounds the 95 GHz DR21W maser. Although none of the clumps of thermal emission are coincident with the 95 GHz DR21W maser position, the peak of the $2_1 \rightarrow 1_1$~E thermal emission line of \meth\ is displaced by only $0.8\arcsec$ from the 95 GHz DR21W maser position, implying the presence of high density gas close to the maser. The 44 GHz Class I \meth\ masers toward DR21W were imaged at high resolution with the VLA (0.2$\arcsec$ half-power beam width) by \citet{kogan1998}. They found two strong masers separated by $\sim 2.8\arcsec$, with several weaker masers located in between, in a linear arrangement that runs from northwest to southeast. The position of the 44 GHz maser at the northwestern end of this linear structure is almost coincident with the 95 GHz maser position given in \citet{pm1990}. 

In this paper, we report the detection of the Zeeman effect in the 44 GHz Class I \meth\ maser line toward DR21W. In Section \ref{sec:obs}, we describe the setup of the observations and the data reduction process. In Section \ref{sec:res}, we present our results, along with a description of the analysis of data for the Zeeman effect. These results are discussed in Section \ref{sec:disc}, and our conclusions are presented in Section \ref{sec:conc}.

\section{Observations and Data Reduction} 
\label{sec:obs}

We observed the $7_{0}-6_1\, A^+$ Class I \meth\ maser emission line at 44\,GHz toward the star forming region \drw\ with the Karl G. Jansky Very Large Array (VLA)\footnote{The National Radio Astronomy Observatory (NRAO) is a facility of the National Science Foundation operated under cooperative agreement by Associated Universities, Inc.} on 2012 April 24 in two consecutive 2\,hr sessions. The VLA was in C-configuration with a maximum baseline of 3.4\,km. The correlator was configured to deliver a single 1\,MHz sub-band with dual polarization products (RR, LL) and 256 spectral channels. The resulting channel spacing was 3.90625\,kHz, which corresponds to a velocity width of 0.0266\,\kmS\ at the observed frequency. In addition to the target source \drw, the calibrator J1331$+$3030 (3C286) was observed to set the absolute flux density scale. The uncertainty in the flux density calibration at the observed frequency, and folding in various observational parameters (e.g., weather, reference pointing, and elevation effects), is expected to be up to 10\%. Table~\ref{tOP}\ gives a summary of our VLA observations.

\begin{deluxetable}{ccrrrrrrrrcrl}
\tablenum{1}
\tablewidth{0pt}
\tablecaption{PARAMETERS FOR VLA OBSERVATIONS
	\protect\label{tOP}}
\tablehead{
\colhead{Parameter} & 
\colhead{Value} }
\startdata
Date &  2012 Apr 24   \\
Configuration & C \\
R.A.~of field center (J2000) & 20$^{\text{h}}$ 38$^{\text{m}}$ 54$\fs$92 \\
Dec.~of field center (J2000) & $+$42$\arcdeg$ 19$\arcmin$ 20$\farcs$6 \\
Total bandwidth (MHz) & 1.0 \\
No.~of channels & 256 \\
Channel spacing (km~s$^{-1}$) & 0.053 \tablenotemark{a} \\
Approx.~time on source (min) & $2 \times 80$ \\
Rest frequency (MHz) & 44069.488 \\
FWHM of synthesized beam & $ 0\, \rlap{\arcsec}.\, 64 \times 0\, \rlap{\arcsec}.\, 50$ \\
& P.A.= $-85.14$\arcdeg \\  \\
Line rms noise (mJy~beam$^{-1}$) \tablenotemark{b} & 8.5 \\
\enddata
\tablenotetext{{\rm a}}{\footnotesize{Image cubes were made by averaging every two channels.}}
\tablenotetext{{\rm b}}{\footnotesize{The line rms noise was measured from the Stokes $I$ image cube using maser line-free channels.}}
\end{deluxetable}

\begin{deluxetable*}{ccccccccccrl}
\tablenum{2}
\tablewidth{0pt}
\tablecaption{FITTED AND DERIVED PARAMETERS FOR DR21W MASERS
	\protect\label{tGauss}}
\tablehead{
\colhead{Maser} &
\colhead{Component} &
\colhead{Intensity} & 
\colhead{Center Velocity\tablenotemark{a}} &
\colhead{Velocity Linewidth\tablenotemark{b}} &
\colhead{Fit Parameter, $b$\tablenotemark{c}} \\
\colhead{} &
\colhead{} &
\colhead{(Jy~beam$^{-1}$)}  &
\colhead{(km~s$^{-1}$)} &
\colhead{(km~s$^{-1}$)} &
\colhead{Hz}}
\startdata
A &1 & $93.03\pm0.48$  & $-2.103\pm0.001$ & $0.344\pm0.002$ & $-23.4\pm3.2$ \\ 
& 2 & \phantom{1}$9.95\pm0.41$ & $-1.227\pm0.033$ & $0.609\pm0.064$ & \nodata \\ 
& 3 & \phantom{1}$4.70\pm1.38$ & $-1.604\pm0.019$ & $0.255\pm0.064$ & \nodata \\ \hline
B & 1 & \phantom{1}$103.73\pm3.63$ & $-2.218\pm0.006$ & $0.263\pm0.005$ & $\phantom{-}13.3\pm2.3$ \\
& 2 & \phantom{1}$73.26\pm5.29$ & $-2.052\pm0.005$ & $0.223\pm0.004$ & $-10.5\pm3.1$
\enddata
\tablenotetext{{\rm a}}{ The center velocity values are with respect to the LSR.}
\tablenotetext{{\rm b}}{ The velocity linewidth was measured at full width at half maximum (FWHM).}
\tablenotetext{{\rm c}}{ The fit parameter, $b=z\Blos$ (see equation~\ref{e1}\ and associated description).}
\end{deluxetable*}

Calibration, deconvolution and imaging, were performed using the Astronomical Image Processing System (AIPS; \citealt{greisen+2003}). The spectral line data of \drw\ were Doppler corrected, and the frequency channel with the brightest maser emission signal was split off, and self-calibrated first in phase, then in both phase and amplitude, and imaged in a succession of iterative cycles. The final self-calibration solutions were applied to the full spectral-line data set of \drw. Final Stokes $I$ and $V$ image cubes were made by averaging every two channels to improve the signal-to-noise; the velocity width in these cubes was 0.053\,\kmS, and the synthesized beamwidth was $0\, \rlap{\arcsec}.\, 64 \times 0\, \rlap{\arcsec}.\, 50$ at full width half maximum (FWHM) and at a position angle of $-85.14^{\circ}$. Note that AIPS calculates the Stokes parameter $I$ as half the sum of the right circular polarization (RCP) and left circular polarization (LCP), so that $I=$~(RCP + LCP)/2, whereas Stokes $V$ is calculated by AIPS as half the difference between RCP and LCP, so that $V=$~(RCP~$-$~LCP)/2; henceforth, all values of $I$ and $V$ are based on this implementation in AIPS. Also note that RCP is defined here in the standard radio convention, in which it is the clockwise rotation of the electric vector when viewed along the direction of wave propagation. 

\section{Results}
\label{sec:res}

\begin{figure}[h!]
\epsscale{1.0}
\plotone{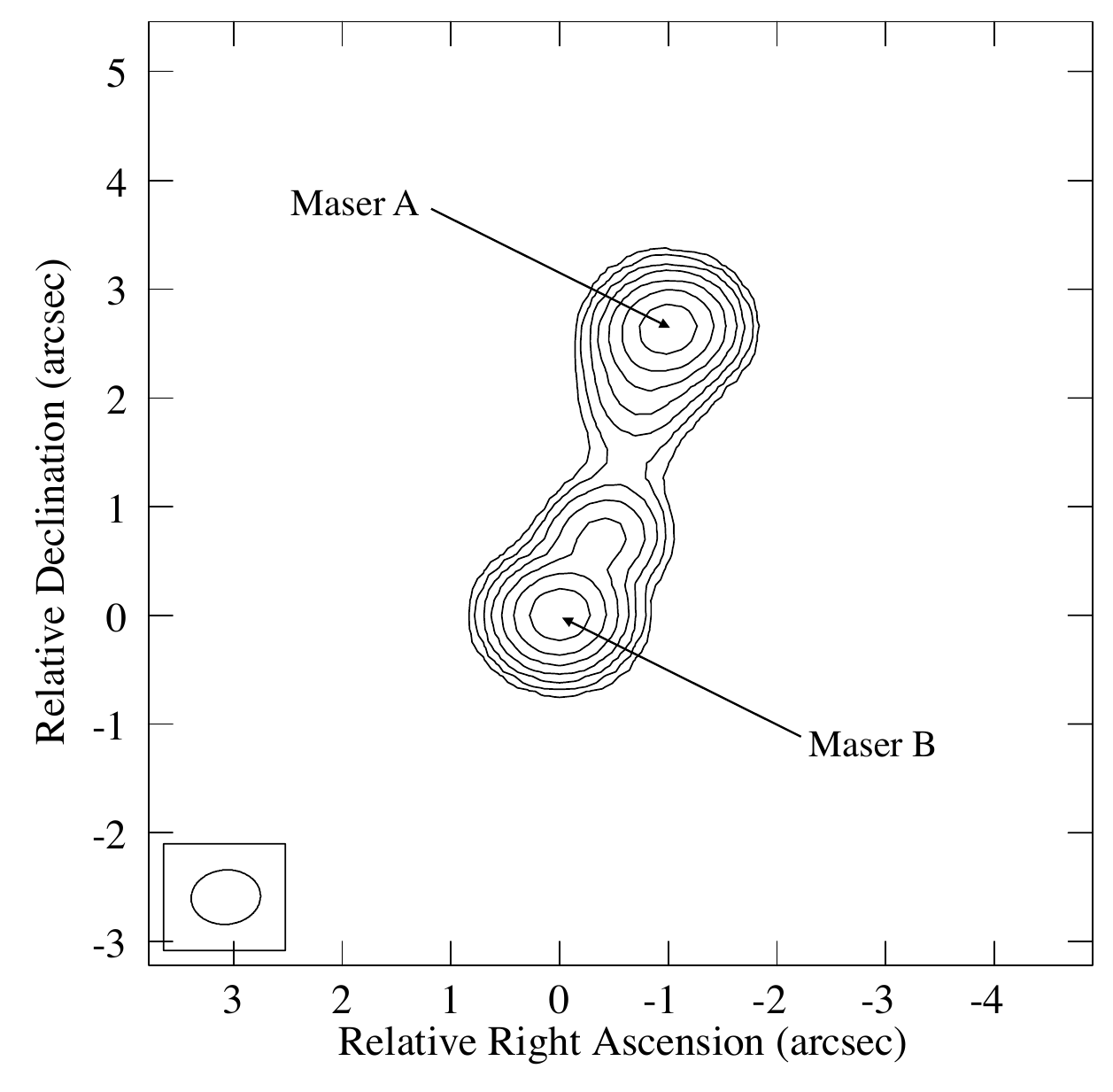}
\caption{Velocity-integrated image of the 44 GHz Class I \meth\ masers in DR21W, showing the masers A and B toward which we report the detection of the Zeeman effect. The velocity range is $-0.3$ to $-3.0$ km s$^{-1}$.  Contours are at $(1, 2, 4, 8, 16, 32, 64) \times 0.42$ Jy beam$^{-1}$ km s$^{-1}$. The position (0,0) in the figure corresponds to $\alpha$=20$^{\text{h}}$ 38$^{\text{m}}$ 54$\fs$9, $\delta=+$42$\arcdeg$ 19$\arcmin$ 20$\arcsec$ (J2000). A higher angular resolution image with a better view of the weaker masers located between A and B can be found in \citet{kogan1998}. \label{fig1}}
\end{figure}

\begin{figure}[htb!]
\plotone{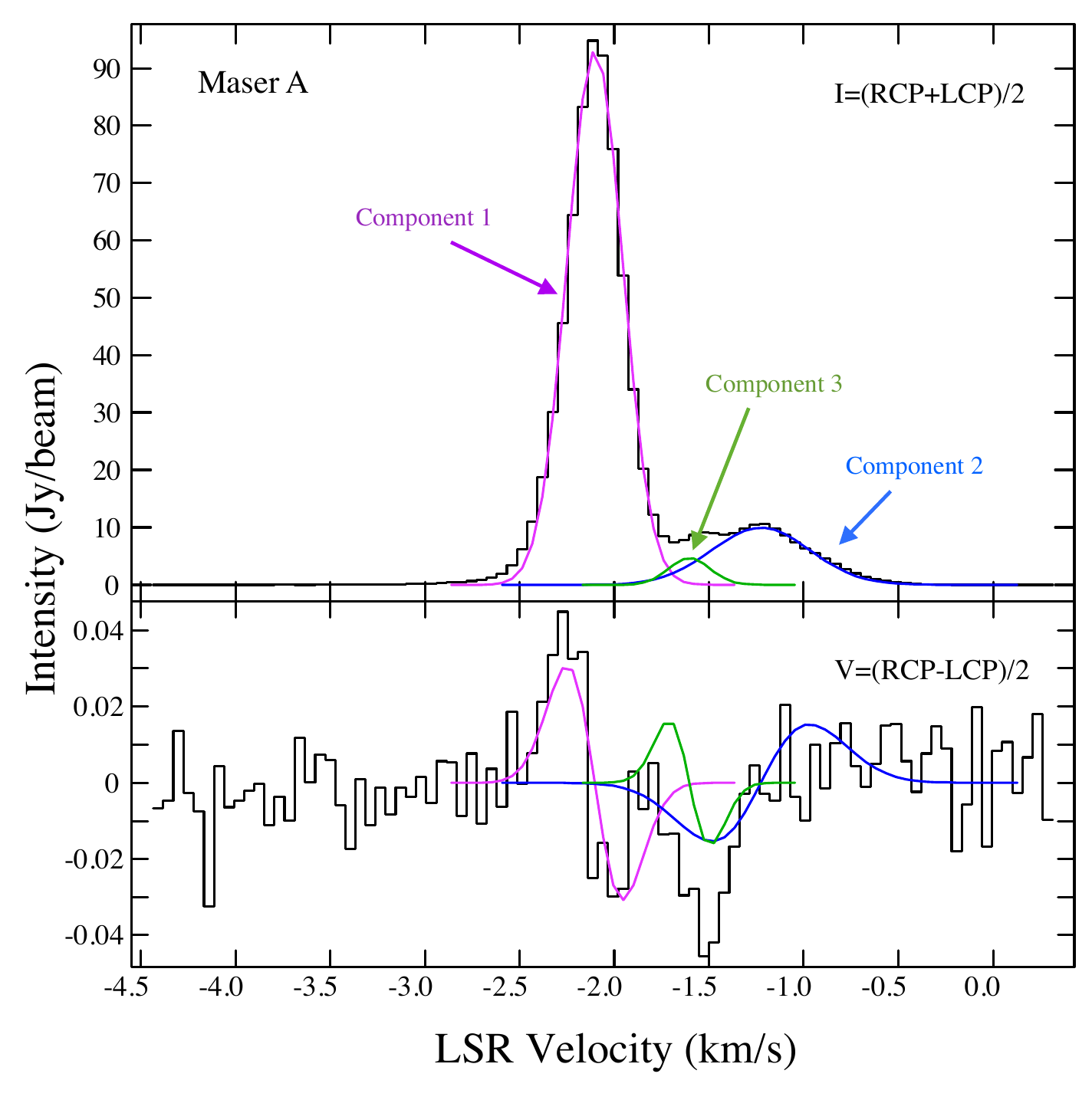}
\caption{Stokes $I$ (\textit{upper panel---black histogram-like line}) and Stokes $V$ (\textit{lower panel---black histogram-like line}) profiles toward the maser in DR21W listed as A in Table \ref{tGauss}. The magenta, blue, and green curves in the upper panel show the Gaussian components that we fitted to the Stokes $I$ profile (components 1, 2, and 3 in Table \ref{tGauss}, respectively; these are also marked in the figure). The magenta, blue, and green curves in the lower panel are the derivatives of the corresponding colored curves in the upper panel, scaled by the fitted value of $z\Blos$ for each curve, obtained from our fitting procedure  described in Section~\ref{sec:res}. \label{fig2}}
\end{figure}

\begin{figure}[h!]
\plotone{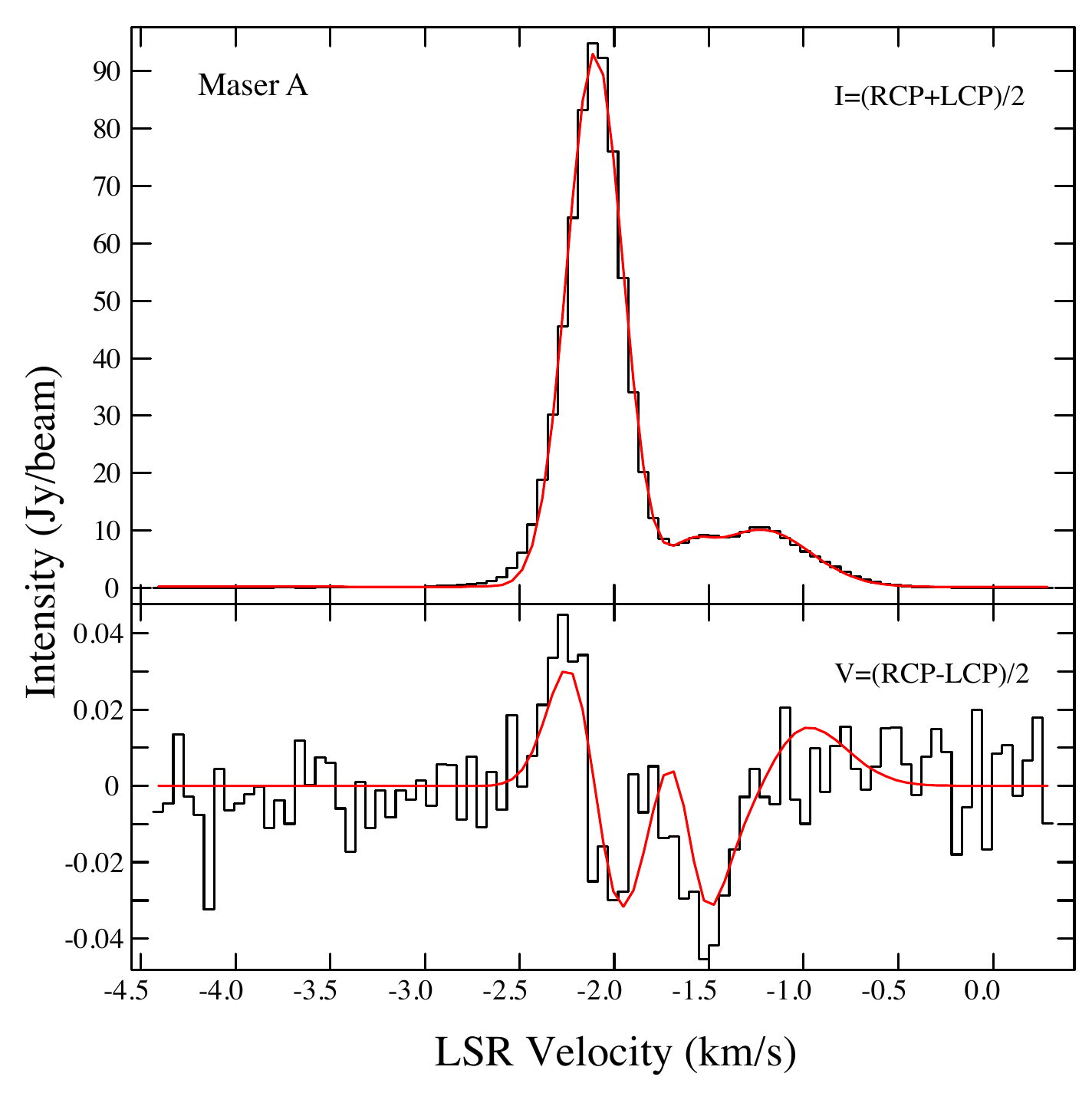}
\caption{Stokes $I$ (\textit{upper panel---black histogram-like line}) and Stokes $V$ (\textit{lower panel---black histogram-like line}) profiles toward the maser in DR21W listed as A in Table \ref{tGauss}. The red curve in the upper panel is the sum of the three Gaussian components shown by magenta, blue, and green curves in the upper panel of Figure~\ref{fig2} (and listed in Table~\ref{tGauss}) that we fitted to the Stokes $I$ profile. The red curve superposed on the Stokes $V$ profile in the lower panel is the sum of the magenta, blue, and green curves shown in the lower panel of Figure~\ref{fig2}; that is, it is the sum of the scaled derivatives of the Gaussian components fitted to the Stokes $I$ profile, where each of the three derivative profiles has been scaled appropriately by the fitted value of $z\Blos$, as described in the caption to Figure~\ref{fig2}. \label{fig3}}
\end{figure}

\begin{figure}[htb!]
\plotone{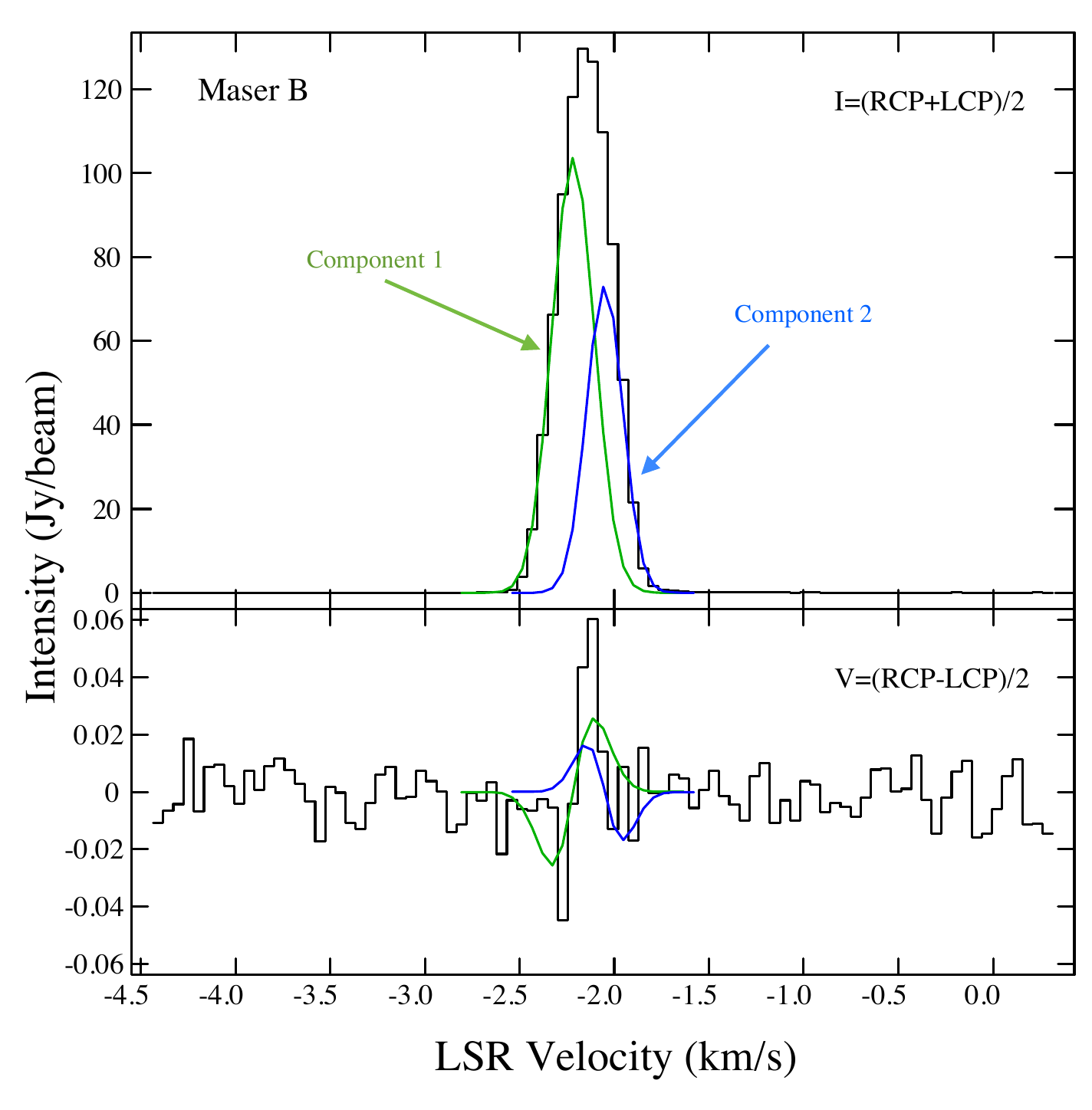}
\caption{Stokes $I$ (\textit{upper panel---black histogram-like line}) and Stokes $V$ (\textit{lower panel---black histogram-like line}) profiles toward the maser in DR21W listed as B in Table \ref{tGauss}. The green and blue curves in the upper panel show the Gaussian components that we fitted to the Stokes $I$ profile (components 1 and 2 for maser B in Table \ref{tGauss}, respectively). The green and blue curves in the lower panel are the derivatives of the corresponding colored curves in the upper panel, scaled by the fitted value of $z\Blos$ for each curve, obtained from our fitting procedure  described in Section~\ref{sec:res}. \label{fig4}}
\end{figure}

\begin{figure}[htb!]
\plotone{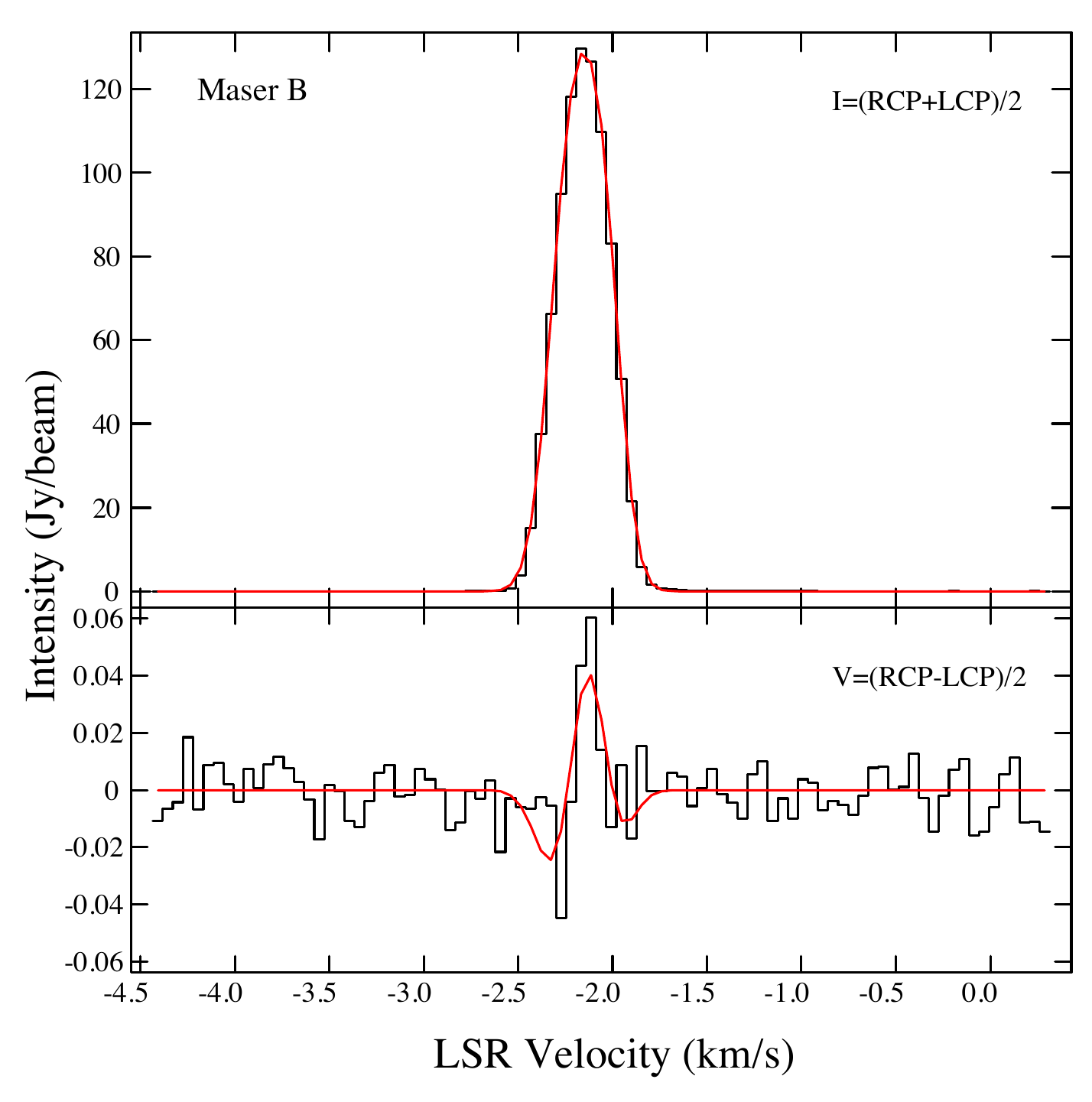}
\caption{Stokes $I$ (\textit{upper panel---black histogram-like line}) and Stokes $V$ (\textit{lower panel---black histogram-like line}) profiles toward the maser in DR21W listed as B in Table \ref{tGauss}. The red curve in the upper panel is the sum of the green and blue Gaussian components shown in Figure~\ref{fig4} (and listed in Table~\ref{tGauss}) that we fitted to the Stokes $I$ profile. The red curve superposed on the Stokes $V$ profile in the lower panel is the sum of the two colored curves shown in the lower panel of Figure~\ref{fig4}; that is, it is the sum of the scaled derivatives of the Gaussian components fitted to the Stokes $I$ profile, where each of the two derivative profiles has been scaled appropriately by the fitted value of $z\Blos$, as described in the caption to Figure~\ref{fig4}. \label{fig5}}
\end{figure}

We observed two prominent maser sources in the 44 GHz Class I \meth\ maser line toward DR21W. These two sources are arranged at the northwestern and southeastern ends respectively of a $\sim$3$\arcsec$ linear arrangement with weaker masers in between (Figure~\ref{fig1}). Figure~\ref{fig2} shows the Stokes $I$ and $V$ profiles toward the 44 GHz Class I \meth\ maser at the northwestern end of this linear arrangement; henceforth, we will designate this as maser A. We fitted three Gaussian components to the Stokes $I$ profile of this maser; the intensity, velocity at line center and linewidth of each component are listed in Table~\ref{tGauss}. These three individual Gaussian components are also shown in the upper panel of Figure~\ref{fig2} by magenta, blue, and green curves, and their composite profile is shown together with the observed Stokes $I$ profile in the upper panel in Figure~\ref{fig3}. The strongest component in maser A with an intensity of 93~Jy~beam$^{-1}$ is at a center velocity of $-2.1$~\kmS\ with respect to the LSR. The other two components are lower in intensity by a factor of almost 10, or greater (9.95 and 4.70 Jy~beam$^{-1}$ respectively), and are at LSR velocities of $-1.2$~\kmS\ and $-1.6$~\kmS, respectively. The fitted components are quite narrow, with velocity linewidths varying from 0.26~\kmS\ to 0.61~\kmS. For the other prominent maser in the southeast which we will henceforth designate as maser B, the Stokes $I$ and $V$ profiles are shown in Figure~\ref{fig4}. For this maser, we fitted two Gaussian components to the Stokes $I$ profile, and their intensity, velocity at line center, and linewidth are also given in Table~\ref{tGauss}. The two Gaussian components fitted to Stokes $I$ are shown by green and blue profiles in the upper panel of Figure~\ref{fig4}, and their composite profile is shown together with the observed Stokes $I$ profile of maser B in the upper panel in Figure~\ref{fig5}. The two fitted components are quite similar in intensity, with the stronger one being about 1.4 times the other. Their velocities at line center are separated by only 0.17~\kmS, and both components are very narrow, with velocity linewidths comparable to the fitted component in maser A that has the least velocity width.  

To determine the magnetic field strength, we fitted a numerical frequency derivative of the Stokes $I$ spectrum to the Stokes $V$ spectrum. This is now standard procedure, and has been described in detail in many sources; see, e.g., \citet{ms2017}. The key point is that the Stokes $V$ profile is fit simultaneously to the derivative of the $I$ profile and a scaled replica of the $I$ profile itself via the equation (\citealt{th1982}; \citealt{sault1990}):
\begin{equation}
V = aI + \frac{b}{2}\, \frac{dI}{d\nu}  \label{e1}
\end{equation}
The scaled replica of the $I$ spectrum is included in the fit to account for small calibration errors in RCP versus LCP; for all results reported in this paper, $a \lesssim 10^{-3}$. The information about the magnetic field is contained in the other fit parameter, $b$, which is equal to $z \Blos$, where $z$ is the Zeeman splitting factor for the molecule being observed, and \Blos~is the line of sight magnetic field strength. Our practice has been to present results for $z\Blos$ in Hz because values of $z$ for 44 GHz methanol were not available. Recently, however, \citet{lankhaar2018} published the results of quantum mechanical calculations of $z$ for a wide array of methanol maser lines, including the 44 GHz Class I \meth~maser line. 

Our procedure to fit the derivative of Stokes $I$ to $V$ using equation~(\ref{e1}) was carried out using the task ZEMAN in AIPS (\citealt{greisen2015}). The key advantage of this task is that it allows multiple Gaussian components in $I$ to be fitted simultaneously to $V$, with each Gaussian component fitted for a different $b$, and hence a different magnetic field value. Thus the magenta curve in the lower panel of Figure~\ref{fig2} is the derivative of the magenta curve in the upper panel of that figure, scaled by the fitted value of $b$, and likewise for the blue and green curves respectively. The sum of these three curves is the red curve in the lower panel of Figure~\ref{fig3}, and it matches well the black histogram-like line of Stokes $V$, also shown in the lower panel of that figure. To claim a significant detection, we require that the ratio of fitted $b$ to the formal error in the fit be at the 3-$\sigma$ level or higher. For maser A, this criterion is satisfied only by the strongest maser, for which $b=-23.4\pm3.2$~Hz. If we use $z=-0.920$~Hz $\mu$G$^{-1}$ from \citet{lankhaar2018} for the $F=5 \rightarrow 4$ hyperfine in the 44 GHz \meth\ transition, we obtain a value of $|\Blos|=25.4$~mG for maser A (but see the discussion in Section~\ref{sec:disc}). For maser B, however, matters are complicated. Certainly, both the fitted Gaussian components for maser B show a formally significant detection, with $b=13.3\pm2.3$~Hz for the stronger component (component 1 in Table~\ref{tGauss}), and $b=-10.5\pm3.1$~Hz for the other (component 2). However, the Stokes $V$ profile for maser B has structure over only a few velocity channels. We discuss this in more detail in the next section.

\section{Discussion}
\label{sec:disc}

Masers A and B in our observations are located at the northwestern and southeastern ends respectively of a linear arrangement of masers spanning about 3$\arcsec$, with weaker masers in between (Figure~\ref{fig1}). This spatial arrangement is consistent with the high resolution observations (VLA A-configuration, 0.2$\arcsec$ half-power beam width) of \citet{kogan1998}. We note that our masers are at velocities with respect to the LSR of $-2.1$~\kmS\ and $-2.2$~\kmS\, whereas \citet{kogan1998}\ list theirs at $-2.3$~\kmS\ and $-2.5$~\kmS\ respectively; however, we don't believe this is an issue since their velocity resolution is 0.17~\kmS\ and ours is 0.053~\kmS. Our center velocity of $-$2.1~\kmS\ matches that reported by \citet{pratap2008} who observed the 44 GHz \meth\ maser emission with the Haystack 37~m telescope with a velocity resolution of 0.014~\kmS.

Extracting information about the magnetic field from the fit parameter $b$ in equation~(\ref{e1}) requires knowledge of the Zeeman splitting factor $z$, since $b=z\Blos$. Until recently, results of Zeeman effect observations in \meth\ masers had been limited to publishing the value of $b$, since $z$ for the 44 GHz Class I methanol maser line was not known. Recently, \citet{lankhaar2018} reported the Zeeman splitting factors for the most prominent methanol lines. For the 44 GHz \meth\ maser, they published values of $z$ for each of its eight hyperfine transitions; however, they assumed that the $F=5\rightarrow4$ hyperfine transition is favored, and that the 44 GHz \meth\ maser line is caused by this transition. If we use their value of $-$0.920~Hz~mG$^{-1}$ for the $F=5\rightarrow4$ hyperfine transition of the 44 GHz \meth\ maser line, we obtain a value of 25~mG for the line-of-sight magnetic field traced by maser A. However, this magnetic field value should be considered as a lower limit, since the $z$ values of the other hyperfine transitions yield larger magnetic field values. Another possible detection in this source is that for maser B. The 104 Jy Gaussian component in maser B yields $b=13.3\pm2.3$~Hz, whereas the 73~Jy Gaussian component yields $b=-10.5\pm3.1$~Hz. Formally, both of these are significant detections based on our criteria that the ratio of fitted $b$ to the formal error must be at the 3-$\sigma$ level or higher. However, we will only claim them as tentative detections for now, since the Stokes $V$ profile for maser B has structure over only a limited range of channels that is narrower than the velocity extent of the maser. If the structure in $V$ is truly caused by the magnetic field in the regions traced by the two components of maser B, then using the \citet{lankhaar2018} value of $-$0.920~Hz~mG$^{-1}$ would give $-$14.5~mG and 11.4~mG respectively for the line-of-sight magnetic field traced by these two maser components. The magnitude of both fields is certainly within the expected range of values in such regions. Another point of interest is that the signs of the fields are opposite. Zeeman observers typically interpret this as a field reversal between the two regions traced by these components. However, \citet{lankhaar2018} have proposed an alternative scenario in which they ascribe the change in sign of the line-of-sight field to the masing transition itself being caused by a different hyperfine component than the one for which $z=-0.920$~Hz~mG$^{-1}$. If we follow their procedure and assume that the hyperfine with $z=0.205$~Hz~mG$^{-1}$ is favored for the oppositely polarized maser, we obtain $\Blos =64.9$~mG in the region traced by the stronger Gaussian component 1 of the maser B. This is certainly on the higher side of the range of magnetic fields expected in regions traced by Class I methanol masers, but it is nevertheless plausible. Class I methanol masers are known to form in outflows in star forming regions. Since the magnetic field will be enhanced by shock compression in such outflows, values for the magnetic field in the tens of milliGauss are expected in such regions. Given that the $F=5 \rightarrow 4$ hyperfine yields a lower limit for $B_{\text{los}}$, it is also worth asking if we can rule out some of the other hyperfines for which \citet{lankhaar2018} calculated the Zeeman splitting factors, $z$. For example, if the $F=7\rightarrow6$ hyperfine were responsible for the 44 GHz \meth\ maser, our detected field in maser A would be 60 times higher than 25~mG, or 1500~mG. Fields as large as this are unlikely at the densities in regions where 44 GHz \meth\ masers are formed (\citealt{leurini+2016}). If we arrange the eight hyperfine lines of the 44 GHz \meth\ maser transition in decreasing absolute value of $z$, the first four give plausible values of the field (tens of mG) for maser A; the next two yield $|\Blos| \sim 120$~mG and border on the realm of the possible, and the last two give values of $220$~mG and $1500$~mG respectively, and can be ruled out. Since $z$ for the $F=5 \rightarrow 4$ hyperfine yields the lowest value for $B_{\text{los}}$, similarly to \citet{lankhaar2018}, we assume in the rest of the discussion below that the $F=5 \rightarrow 4$ hyperfine is responsible for the 44 GHz \meth\ maser transition.

Reporting on Zeeman observations invariably involves a discussion of whether the detected signal could be due to instrumental effects, or if it could be caused by processes other than the Zeeman effect. First we consider the instrumental effects, in which a velocity gradient across an extended source could cause structures in Stokes $V$ that resemble those due to the Zeeman effect. Fortunately, masers are point sources confined to a narrow velocity range, and therefore it appears unlikely that a velocity gradient could cause a Zeeman pattern in masers. Moreover, such a false pattern would depend on parallactic angle. Since our observations were carried out in two sessions that span different parallactic angles, the consistency of our Stokes $V$ profile between these two sessions gives us confidence that our detection is not a spurious instrumental effect. There are, of course, other processes that could contribute to structure in the Stokes $V$ profile. \citet{wiebe1998} found that in masers with strong linear polarization, changes in the orientation of the magnetic field along the line of sight could cause rotation of the linear polarization vector to produce circular polarization. Observing with the Korean VLBI network telescopes in single dish mode, \citet{kang+2016} measured $(2.0 \pm 0.2)\, \%$ linear polarization in the 44 GHz \meth\ masers toward DR21W. Given this low value, it is unlikely that linear polarization is causing the observed Stokes $V$ in our observations. Another possible non-Zeeman origin for Stokes $V$ comes from \citet{houde2014}, who found that maser radiation scattering off foreground molecules can enhance antisymmetric spectral profiles in SiO masers. Thus, if the Stokes $V$ profile were deemed to be entirely due to the Zeeman effect, one would obtain a much larger value for the magnetic field traced by these SiO masers. This seems unlikely for our methanol maser observations, since fields like 25~mG detected in maser A are expected in such regions. Finally, as described in \citet{vlemmings2011}, a rotation of the axis of symmetry for the molecular quantum states could also cause circular polarization if the maser stimulated emission rate $R$ becomes larger than $g\Omega$, the frequency shift due to the Zeeman effect. In our data for DR21W, $g\Omega \approx 25$~s$^{-1}$. The stimulated emission rate $R$ is given by
\begin{equation} R \simeq \frac{AkT_b \Delta \Omega}{4\pi h \nu}  \label{e.2} \end{equation}
(\citealt{vlemmings2011}), where $A=25.913 \times 10^{-8}$~s$^{-1}$ is the Einstein coefficient for the $F=5\rightarrow4$ hyperfine transition of the 44 GHz \meth\ maser line (\citealt{lankhaar2018}), $T_b$ is the maser brightness temperature, and $\Delta \Omega$ is the maser beaming angle. For the $\nu=44069.488$~MHz methanol maser in which we detected the Zeeman effect in DR21W, we find that $T_b \simeq 3 \times 10^6$~K, and based on \citet{slysh2009}, we calculate $\Delta \Omega \simeq 0.03$. Using these values, we find from equation~(\ref{e.2}) that $R \approx 10^{-3}$~s$^{-1}$. Therefore, $R \ll g\Omega$, implying that it is unlikely that a rotation of the axis of symmetry for the molecular quantum states is causing the observed    splitting responsible for the shape of the Stokes $V$ profile. Moreover, such a rotation would cause an intensity-dependent polarization, but the component with higher intensity in maser B shows a smaller magnetic field than that detected in maser A.

Since Class I \meth\ masers occur in outflows, the shocks in these regions likely amplify the magnetic field in proportion to the gas density, so that
\begin{equation} \frac{B_{\text{post}}}{B_{\text{pre}}} 
	= \frac{n_{\text{post}}}{n_{\text{pre}}}  \label{e.3} \end{equation}
where post and pre refer to the post-shocked and pre-shock regions respectively (\citealt{sarma2008}). We can use equation~(\ref{e.3}) to calculate the magnetic field in the pre-shock gas. \citet{leurini+2016}\ found that bright Class I \meth\ masers occur in high density regions with $n(\text{H}_2) \sim 10^{7-8}$~cm$^{-3}$, so we will use this as our value for $n_{\text{post}}$. For the particle density in the pre-shock gas, we can use the observations of \citet{lw1997}, who found the column density from thermal methanol emission in DR21W to be $N$(\meth)~$=1.0 \times 10^{15}$~cm$^{-2}$. Using their estimate of $2.2 \times 10^{-8}$ for the methanol abundance gives a column density for hydrogen, $N (\text{H}_2) = 4.5 \times 10^{22}$~cm$^{-2}$. This agrees well with the value of $N (\text{H}_2) = 3.7 \times 10^{22}$~cm$^{-2}$ determined by \citet{jakob2007} from dust continuum observations toward DR21W. If we use an average source size of $6\arcsec$, based on \citet{lw1997}, as the dimension along the line of sight, our derived value of $N(\text{H}_2)$ gives a particle density of $n(\text{H}_2)=3.3 \times 10^{5}$~cm$^{-3}$, which we will use as our value of $n_{\text{pre}}$. With these values, the ratio $n_{\text{post}}/n_{\text{pre}} = 30$-300. With the line-of-sight magnetic field $(B_{\text{post}})_{\text{los}}=25$~mG from our Zeeman detection in maser A, and equation~(\ref{e.3}), we calculate the line-of-sight magnetic field in the pre-shock region and find it to be  $(B_{\text{pre}})_{\text{los}}=0.8$~mG if the post-shock particle density is $n_{\text{post}}=10^7$~cm$^{-3}$, or 0.1~mG if $n_{\text{post}}=10^8$~cm$^{-3}$. Now, while there are no thermal line measurements of the Zeeman effect in DR21W, \citet{falga2008} measured line-of-sight magnetic fields of 0.36 mG and 0.71 mG in the sources MM1 and MM2 about $3.5\arcmin$ away in the nearby region DR21(OH) using the Zeeman effect in the thermal CN line. They measured these fields in gas of particle density $n (\text{H}_2) \sim 1.7 \times 10^5$~cm$^{-3}$, similar to the pre-shock density of $3.3 \times 10^5$~cm$^{-3}$ that we've used in our calculations above for DR21W. Both these magnetic field values detected in the thermal CN line about $3.5\arcmin$ away in DR21(OH) compare well with the predicted range of 0.1-0.8~mG for the line-of-sight pre-shock field in DR21W based on our observations. 

Finally, we can use our detected value for $\Blos$ in DR21W to compare the magnetic energy density in the post-shocked regions to the kinetic energy density. The magnetic energy density is given by $B^2/8\pi$, where $B^2 = 3\Blos^2$ (\citealt{crutcher1999}). For $\Blos = 25$~mG from our observations, the magnetic energy density is then $7.5 \times 10^{-5}$~erg cm$^{-3}$. If $\Blos$ is larger due to a different hyperfine transition being responsible for the 44 GHz \meth\ maser, the magnetic energy density will be even larger. Next, the kinetic energy density, including the contribution of both thermal and turbulent motions, is given by $(3/2)\, m n \sigma^2$. In this expression, the mass $m = 2.8\, m_p$, where $m_p$ is the proton mass, and the numerical factor of 2.8 also accounts for 10\% He, and the velocity dispersion, $\sigma = \Delta v/(8 \ln\, 2)^{1/2}$. To get a true picture of the kinetic energy density in gas of $n \sim 10^{8}$~cm$^{-3}$, one needs to use a larger value of $\Delta v$ than the 0.2-0.6~\kmS\ for the masers in our observations (see Table~\ref{tGauss}). From observations of thermal methanol emission toward DR21W, \citet{lw1997}\ found $\Delta v \sim 3.5$~\kmS, and using this value of $\Delta v$ to calculate $\sigma$, we find that the kinetic energy density in DR21W is $\sim 1.5 \times 10^{-5}$~erg~cm$^{-3}$. Thus, the magnetic energy density in the post-shocked gas is at least comparable to, or higher than, the kinetic energy density, and we would expect the magnetic field to play a significant role in shaping the dynamics in these post-shocked regions of DR21W.

\section{Conclusions}
\label{sec:conc}

We have detected the Zeeman effect in the Class I \meth\ maser line toward the high mass star forming complex DR21W. For maser A, which lies at the northwestern end of a linear arrangement of two prominent masers with weaker masers in between (Figure~\ref{fig1}), we fitted three Gaussian components (Table~\ref{tGauss}, and Figure~\ref{fig2}). In the strongest component for maser A, we have a significant detection of the Zeeman effect and we find that $z\Blos=-23.4\pm3.2$~Hz. If we assume, following \cite{lankhaar2018}, that the $F=5\rightarrow4$ hyperfine is responsible for the 44 GHz \meth\ maser, our detected value of $z\Blos$ corresponds to a magnetic field of magnitude $|\Blos|=25$~mG; the field could be higher if one of the other seven hyperfine transitions is responsible for the 44 GHz \meth\ maser, but we have ruled out some of these hyperfines since fields of the order of hundreds of mG are unlikely in these regions. In maser B, which is at the southeastern end of the linear arrangement, we fitted two Gaussian components. In both these components, we also have formally significant detections of $z\Blos=13.3\pm2.3$~Hz and $z\Blos=-10.5\pm3.1$~Hz, corresponding to $\Blos$ values of magnitude 14.5~mG and 11.4~mG respectively; again, these values could be higher if a different hyperfine transition is responsible for the 44 GHz \meth\ maser. However, we consider the result for maser B only as a tentative detection, since the Stokes $V$ profile for maser B has structure over only a limited range of channels that is narrower than the velocity extent of the maser. Since Class I masers are formed in outflows where the shock will compress the magnetic field in proportion to the gas density, we can use our detected value of $\Blos=25$~mG toward maser A (derived assuming the $F=5 \rightarrow 4$ hyperfine as the favored transition), together with particle densities obtained from the literature, to predict that the line-of-sight magnetic field in the pre-shock gas should be $\sim$0.1-0.8 mG. Although there are no Zeeman detections in thermal lines toward DR21W for a direct comparison, the range of values predicted for the line-of-sight pre-shock magnetic field is in good agreement with fields of 0.36 mG and 0.71~mG in gas of comparable density $n \sim 1.7 \times 10^5$~cm$^{-3}$,  measured via the Zeeman effect in the thermal CN line about $3.5\arcmin$ away in DR21(OH). We have also determined that magnetic fields likely play a significant role in the dynamics of the post-shocked gas in this region.

\acknowledgments

\facility{VLA.}

\end{document}